# Determining the Number of Non-Spurious Arcs in a Learned DAG Model: Investigation of a Bayesian and a Frequentist Approach

**Jennifer Listgarten and David Heckerman**
Microsoft Research
One Microsoft Way
Redmond, WA 98052

## Abstract

In many application domains, such as computational biology, the goal of graphical model structure learning is to uncover discrete relationships between entities. For example, in our problem of interest concerning HIV vaccine design, we want to infer which HIV peptides interact with which immune system molecules (HLA molecules). For problems of this nature, we are interested in determining the number of non-spurious arcs in a learned graphical model. We describe both a Bayesian and frequentist approach to this problem. In the Bayesian approach, we use the posterior distribution over model structures to compute the expected number of true arcs in a learned model. In the frequentist approach, we develop a method based on the concept of the *False Discovery Rate*. On synthetic data sets generated from models similar to the ones learned, we find that both the Bayesian and frequentist approaches yield accurate estimates of the number of non-spurious arcs. In addition, we speculate that the frequentist approach, which is non-parametric, may outperform the parametric Bayesian approach in situations where the models learned are less representative of the data. Finally, we apply the frequentist approach to our problem of HIV vaccine design.

## 1 Introduction

In many application areas where graphical models are used and where their structure is learned from data, the end goal is neither prediction nor density estimation. Rather, it is the uncovering of discrete relationships between entities. For example, in computational biology, one may be interested in discovering which proteins within a large set of proteins interact with one another or which miRNA molecules target which mRNA molecules. In these problems, relationships can be represented by arcs in a graphical model. Consequently, given a learned model, we are interested in knowing how many of the arcs are real or *non-spurious*.

**Modeling of HIV Data**

The modeling problem which motivates our work is related to the rational design of HIV vaccines. As a bit of background, there are two arms of the adaptive immune system: the humoral arm, which manufactures antibodies, and the cellular arm, which recognizes cells that are infected and kills them. Our HIV vaccine design concentrates on the cellular arm. The cellular arm kills infected cells by recognizing short (8-11 amino-acid long) bits of proteins, known as *epitopes*, that are presented on the surface of most human cells. The epitopes are the result of digestion of proteins (both normal and foreign) within the cell and are presented on HLA (Human Leukocyte Antigen) molecules, which form a complex with the epitope before it is presented. Special cells of the immune system, known as T-cells, recognize the epitope–HLA complexes, and kill the cell if the epitopes correspond to foreign or non-self proteins.

There are hundreds of types of HLA molecules across the human population; each person has from three to six different types. Furthermore, the epitopes presented by one HLA type are typically different from those presented by another type. (This diversity is thought to be useful in preventing a single virus or other pathogen from destroying the human race.) One possible design of a cellular vaccine would then be to identify a set of peptides that are epitopes for a large number of people and assemble them in a delivery mechanism that would train T-cells to recognize and kills cells infected by HIV.

To identify peptipe-HLA pairs that are epitopes, one can take a peptide and mix it with the blood (which includes T-cells) of an individual and watch for the release of gamma interferon, which indicates that the T-cell killing mechanism has been activated—an *ELIspot* assay [e.g., Goulder et al., 2001]. If the reaction occurs, it is likely that the peptide is an epitope for one of the HLA types of the patient. Although it is impossible to discern the HLA type or types that are responsible for a reaction from a single patient, with data from many patients we can infer (probabilistically) which of the patients HLA types are causing the observed immune reactions.

To perform this inference, we model our data using a bi-partite graph with noisy-OR distributions as shown in Figure 1. Recall that a noisy-OR distribution is based on the assumption of *independence of causal influence* [2] and takes the form,

$$p(y_j^k = 0|\{q_{ij}\}, q_0) = (1 - q_0) \prod_{\{i|h_i^k=1\}} (1 - q_{ij})$$
$$p(y_j^k = 1|\{q_{ij}\}, q_0) = 1 - p(y_j^k = 0|\{q_{ij}\}, q_0),$$

where $h_i^k = 1$ [0] denotes that patient $k$ has [does not have] HLA allele $i$, and $y_j^k$ denotes the observed, binary reactivity for peptide $j$ in patient $k$, and $q_{ij}$ is the link probability that peptide $j$ reacts with HLA $i$ (that is, $q_{ij}$ is the probability that HLA $i$ causes a reaction to peptide $j$ in the absence of any other HLA molecules). We note that the values of $q_{ij}$ are not always equal to one due to (*e.g.*) noise in the assay. We also note that although there are correlations among the HLA variables (due to linkage disequilibrium), we are not interested in these correlations in this application.

Given a learned set of arcs representing peptide–HLA interactions, we wish to determine how many of these arcs are real, so that we can assess whether any of them merit further confirmatory testing. In general, *given a single model structure, learned by any method, we wish to determine its expected number of non-spurious arcs.* This determination allows researchers to generate specific, testable hypotheses and gauge confidence in them before considering whether to pursue them further.

**Two Approaches**

We use the term *arc hypothesis* to denote the event that an arc is present in the underlying distribution of the data. We consider two approaches—one Bayesian and one frequentist—to the problem of estimating the number of true arc hypotheses in a given learned model.

In our Bayesian approach, we compute the expected number of true arc hypotheses given data, where expectation is computed by averaging over all possible graph structures according to their posterior probability. In our frequentist approach, we estimate and control the *False Discovery Rate* (FDR) [3] of a set of arc hypotheses. The FDR is defined as the (expected) proportion of all hypotheses (*e.g.*, arc hypotheses) which we label as true, but which are actually false (*i.e.*, the number of false positives divided by the number of total hypotheses called true).

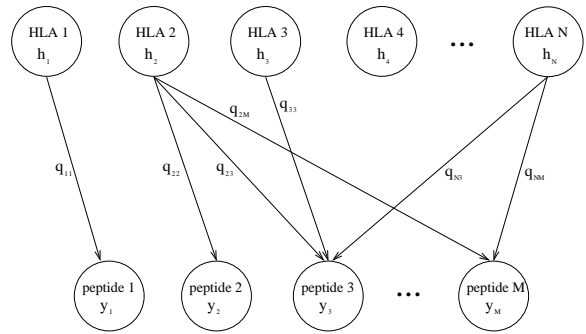

Figure 1: A bi-partite graph used to model which HLA types interact with which HIV peptides. The probability of each peptide having a reaction is parameterized by a noisy-OR distribution over its parents. All nodes are observed, and we are interested in finding which arcs are present. A peptide may have zero or several incoming arcs. An HLA molecule may have zero or several outgoing arcs. Each person has between three and six HLA molecules. Thus, for a given patient, between three and six HLA nodes will be "on."

Both our Bayesian and frequentist approaches address only the number of true arc hypotheses within a given set. They do not place a probability distribution (jointly or marginally) on the arc hypotheses in the set. Although the latter goal can be approached with Bayesian or frequentist methods,[1] we focus on the former goal here.

In our evaluations, we concentrate on directed acyclic graphs (DAGs) for discrete variables with known variable orderings, as our problem of interest has these properties. On synthetic data sets generated from models similar to the ones learned, we shall see that both the Bayesian and frequentist approaches yield accurate estimates of the number of non-spurious arcs. Furthermore, we shall see that our particular frequentist approach is far more computationally efficient than the Bayesian approach. This observation should not be taken as a condemnation of the Bayesian approach, as approximations, such as those based on sampling, may yield accurate results. Nonetheless, we speculate that the frequentist approach, which is non-parametric, may outperform our parametric Bayesian approach in situations where the models learned are less representative of the data.

The remainder of this paper is structured as follows. First, we provide a more detailed account of our Bayesian approach to computing the expected number of true arcs in a given, learned graph structure. Second, we introduce our frequentist method for estimating the FDR of the arcs in a given, learned graph structure. Third, we compare the two methods on CPT-based (conditional probability table) and noisy-OR-based synthetic data sets. Finally, we present results on real data for our problem of HIV vaccine design.

---
[1] Friedman and Koller[4] describe the Bayesian approach. Efron [5] describes a per-hypothesis or "local" FDR that can be applied to this problem.

## 2　A Bayesian Approach

As we have discussed, the Bayesian approach is to compute the expected number of true arc hypotheses in a learned model given data, where expectation is computed by averaging over all possible model structures according to their posterior probability. Let the set of arc hypotheses be those in some learned model, $G^l$, which has been found by any method (e.g., Bayesian structure search). Using $t^l$ to denote the number of true arc hypotheses in $G^l$, and $E_D(t^l)$ to denote the Bayesian expectation of $t^l$ given data $D$, we have

$$E_D(t^l) = \sum_G P(G|D) \, f(G, G^l), \quad (1)$$

where the summation is over all possible graphical model structures, $G$, and where $f(G, G^l)$ is the number of arcs that appear in both $G$ and $G^l$. The posterior $P(G|D)$ is computed by $P(G|D) \propto P(G) \, P(D|G)$, where $P(G)$ is the prior over graph structures. The quantity $P(D|G)$, known as the *marginal likelihood,* is computed in turn by integrating over the parameters $\theta$ of each model structure $G$. Thus, we obtain

$$P(G|D) \propto P(G) \, P(D|G) = P(G) \int_\theta P(\theta, D|G). \quad (2)$$

For models with CPT-based parameterizations and independent Dirichlet parameter priors, $P(D|G)$ can be computed efficiently in closed form [6]. For many other models such as noisy-OR based models, however, $P(D|G)$ cannot be computed exactly. Consequently, an approximation to it such as the Laplace approximation or the Bayesian Information Criterion (BIC) is used. These approximations are often much slower to compute, especially when they involve the iterative fitting of model parameters.

When the variable ordering is known, there are no missing data, and the parameters for each node are mutually independent, the problem of learning a graphical model structure on $N$ nodes decomposes into $N$ independent problems, where each problem involves learning the parents of a given node. Thus the sum in Equation 1 becomes a sum over all possible parent sets for a single node, where the set of potential parents for each node is induced by the variable ordering. This computation remains exponential [4], but sometimes can be approximated by summing over parents sets up to some maximum size. For our experiments evaluating the Bayesian approach, we use the CPT-based Alarm network [7] which has 37 nodes and no more than four arcs incident on a node. In these experiments, we obtain virtually identical results when limiting the size of parents sets to five or six, suggesting that size limit of five produces a good approximation.

Often when learning structure from data, the term $p(G)$ is set to $p(G) \propto \kappa^M$, where $\kappa$ is greater than zero and $M$ is the number of arcs in the graph [8]. In our experiments, we choose $\kappa$ (and one other hyperparameter of the model to be discussed) so that out-of-sample predictions are most accurate. As we shall see in our experiments, a choice of $\kappa < 1$ (i.e., a non-uniform prior on structure) is optimal. More interesting, we find that the $\kappa$ that optimizes prediction is not the one that yields the most accurate estimate of number of non-spurious arcs.

## 3　A Frequentist Approach

When inferring whether a single hypothesis is true or not, statisticians have traditionally relied on the p-value, which controls the number of false positives (type I errors). However, when testing hundreds or thousands of hypotheses simultaneously, the p-value needs to be corrected to help avoid making conclusions based on chance alone (known as the problem of *multiple hypothesis testing*). A widely used, though conservative correction is the Bonferroni correction, which controls the Family Wise Error Rate (FWER). The FWER is a compound measure of error, defined as the probability of seeing at least one false positive among all hypotheses tested. In light of the conservative nature of methods which control the FWER, the statistics community has recently placed great emphasis on estimating and controlling a different compound measure of error, the *False Discovery Rate* (FDR) [3, 9].

In a typical computation of FDR, we are given a set of hypotheses where each hypothesis, $i$, is assigned a score, $s_i$ (traditionally, a test statistic, or the p-value resulting from such a test statistic). These scores are often assumed to be independent and identically distributed, although there has been much work to relax the assumption of independence [10]. The FDR is computed as a function of a threshold, $t$, on these scores, $FDR = FDR(t)$. For threshold $t$, all hypotheses with $s_i \geq t$ are said to be significant (assuming, without loss of generality, that the higher a score, the more we believe a hypothesis). The FDR at threshold $t$ is then given by

$$FDR(t) = E\left[\frac{F(t)}{S(t)}\right],$$

where $S(t)$ is the number of hypothesis deemed significant at threshold $t$ and $F(t)$ is the number of those hypotheses which are false, and where expectation is taken with respect to the true joint distribution of the variables. When the number of hypotheses is large, as is usually the case, one can take the expectation of the numerator and denominator separately:

$$FDR(t) = E\left[\frac{F(t)}{S(t)}\right] \approx \frac{E\left[F(t)\right]}{E\left[S(t)\right]}.$$

Furthermore, it is often sufficient to use the observed $S(t)$ as an approximation for $E[S(t)]$. Thus the computation of $FDR(t)$ boils down to the computation of $E[F(t)]$. One approximation for this quantity which can be reasonable is $E[F(t)] \approx E_0[F(t)]$, where $E_0$ denotes expectation with respect to the null distribution (the distribution of scores obtained when no hypotheses are truly significant), and it is this approach that we take.[2]

Note that the FDR is closely related to *positive predictive value* (PPV), where $PPV(t) = 1 - \frac{F(t)}{S(t)}$. That is, FDR is 1 minus expected PPV.

Applying this approach to estimating the number of non-spurious arcs in a given (learned) DAG model, we take as input a particular structure search algorithm $a$ (which may have hyperparameters such as $\kappa$ that control the number of arcs learned) and generalize $S(\cdot)$ and $F(\cdot)$ to be functions of $a$. In particular, $S(a)$ is the number of arcs found by $a$ and $F(a)$ is the number of those arcs whose corresponding hypotheses are false. As in the standard FDR approach, we use the approximation $E(S(a)) \approx N(D, a)$, where $N(D, a)$ is the number of arcs found by applying $a$ to the real data $D$. In addition, we estimate $E_0(F(a))$ to be $N(D^q, a)$ averaged over multiple data sets $D^q$, $q = 1, \ldots, Q$, drawn from a null distribution. That is,

$$FDR(a) = E\left[\frac{F(a)}{S(a)}\right] \approx \frac{E[F(a)]}{E[S(a)]}$$
$$\approx \frac{(1 + \sum_{q=1}^{Q} N(D^q, a))/Q}{N(D, a)}.$$

The addition of one to the numerator smooths the estimate of $E_0[F(a)]$ so as to take into account the number of random permutations performed.[3]

In our implementation of this approach, we assume that $a$ has the property that it can be decomposed into independent searches for the parents of each node. Given this assumption, when we learn the parent set of a given node, we create the null distribution for that node by permuting the real data for the corresponding variable. This permutation guarantees that all arc hypotheses are false in the null distribution. The generation of these null distributions is computationally efficient as well as non-parametric, making them applicable to situations where the models learned are less representative of the data. There is, however, one theoretical concern about the use of these null distributions, which we address in the discussion section.

To determine whether our approximations are reasonable in practice, we draw samples from synthetic graphical models, run the algorithm above to compute the FDR, and then use the ground truth generating structure to measure the true FDR. When we do so, as for example shown in Figure 4, we see that our approach can produce reasonably accurate estimates of FDR.

By construction, the emphasis of our FDR approach is on the accuracy of the estimate of the number of false positives, and does not examine the number of false negatives. Whereas this emphasis may seem undesirable, it is common for experimenters to be more interested in how many hypothesized interactions are real, rather than how many were missed, because experimenters will in most cases be using resources to pursue the positive hypotheses, not the negative ones. A similar line of reasoning is mentioned by Friedman *et al.* in [4] and by parishioners of FDR [9]. The Bayesian approach allows us, in principle, to quantify both false positives and false negatives, but we do not pursue this computation here.

## 4 Related Work

Friedman *et al.* compute confidence measures on arcs (and other features) of induced Bayesian networks by using the bootstrap [11, 12] (or parametric bootstrap). By resampling their data, and inducing a Bayesian network for each sampled data set, they count how many times a given arc occurs, and estimate the probability of that arc, $\hat{p}_i$ as the proportion of times it is found across all bootstrap samples. Their confidence measure is not an estimate of the number of non-spurious arcs. For example, applying a pathological search algorithm which systematically adds all arcs to their approach yields the estimate $\hat{p}_i = 1$ for every arc. In fact, their measure is a particular estimate of $E[S(t)]$, the denominator of the FDR.

In a later paper, Friedman and Koller follow a Bayesian approach, using MCMC samples over variable orderings to compute marginal probabilities of arc hypotheses [4]. Although they characterize the performance of the MCMC method, they do not determine whether the exact (or approximated) posterior probabilities are accurate or *calibrated* in the sense that hypotheses labeled—say—0.4 are true 40% of the time.

Recently, Pena and colleagues [13] used an approach somewhat similar to Friedman *et al.*'s bootstrap approach. They use a stochastic, greedy structure search algorithm, and run it numerous (e.g., 1000) times to a local optimum. They then score each arc according to the proportion of times it appeared across all local optima found. Although they provide asymptotic guarantees, we do not expect their ap-proach to yield accurate estimates on finite data.

---

[2]For the more traditional application of FDR (*i.e.*, not as applied to graphical models), Storey and Tibshirani offer a clever method to compute $E[F(t)]$ which is less conservative than using $E[F(t)] \approx E_0[F(t)]$ [9]. However, this approach is not appropriate in the present context.

[3]Without this smoothing, if one performed too few random permutations such that $\sum_q N(D^q, a) = 0$ due to sampling error, then the estimate of $E_0[F(a)]$ and hence $FDR(a)$ would also be zero. We prefer our more conservative estimate, especially as the bias it induces diminishes as the number of permutations increases.

In [14], Schäfer and Strimmer develop a frequentist test for edge inclusion in graphical Gaussian models (GGMs) and provide a reasonable model for the null distribution of this test. They assign a score to each edge based on how much it "hurts" the model when each edge is independently removed (this is assessed in the presence of all other possible edges being included in the model—a one-backward-step search for each edge). They then use these scores to compute the FDR for a given set of edges. Applying a similar approach on our data, we found that many real arc hypotheses got low scores, resulting in inaccurate estimates of FDR.

## 5 Experimental Results

As mentioned, we evaluate both the Bayesian and frequentist estimates of the number of non-spurious arcs using synthetic data. In particular, we draw samples from synthetic graphical models, apply the approaches, and then use the generating structure to measure the true number of non-spurious arcs. We perform this evaluation for two generating models: the (CPT-based) Alarm network, and a model (with noisy-OR distributions) learned from our HIV data. In addition, we apply the frequentist approach to our real HIV data.

**Synthetic CPT-based Alarm Data**

In our first set of experiments, we compared the accuracy of our Bayesian approach to that of our frequentist approach, using data generated from the Alarm network [7], which contains 37 CPT-based nodes and 46 arcs. We arbitrarily resolved the four non-compelled edges in the model by placing LVFailure before History, Anaphylaxis before TPR, PulmEmbolus before PAP, and MinVolSet before VentMach. From this model, we generated three data sets with sample size 100 and three with sample size 1000.

The same set of models (*i.e.*, set of arcs) are used in experiments for both the Bayesian and frequentist methods. These models we learned by greedy structure search starting from the empty graph, where a single arc was added or deleted at each step of the search until the structure score could not be increased. The exact (relative) posterior probability given a BDeu prior [6] was used to guide the search. The hyperparameters of the score ($\kappa$ and $\alpha$, the equivalent sample size in the BDeu prior) were chosen to optimize the likelihood of out of sample data using one independent test set whose sample size was equal to that of the training data. We found $\kappa = 0.01$ and $\alpha = 4$ to be optimal for both sizes of data. To evaluate the methods for a range of models with varying numbers of arcs, we fixed $\alpha = 4$ and varied $\kappa$ across the range $[10^{-4}, 5]$. The delete operator was used only occasionally.

The search method $a$ used in the frequentist approach was the one just described. In addition, we used a second search method identical to the first, except the marginal likelihood was approximated by BIC rather than computed exactly. In this second approach, we again used $\alpha = 4$, but varied $\kappa$ in the range $[10^{-2}, 10^4]$. A wider range was used because the BIC approximation tended to underestimate the exact score. For both experimental conditions, we used 10 permutations of the data ($Q = 10$) when estimating $E_0[F]$.

For the Bayesian approach, although the graph structure is given (from the greedy search), we needed to set values for $\kappa$ and $\alpha$ to compute the (exact) marginal likelihood. Again, we set the values to optimize out-of-sample prediction, but here used predictions based on model averaging rather than model selection. We found that $\kappa = 0.1$ and $\alpha = 4$ were optimal for both data set sizes. As mentioned, we limited the sum in Equation 1 to parents sets of size five or less. Results were nearly identical when the limit was increased to parent sets of size six or less.

The results of the comparison are shown in Figure 2, which plots the expected positive predictive value (as predicted by each of the two approaches), against the actual positive predictive value according to the generating structure of the Alarm network. The expected PPVs (positive predictive values) plotted for the frequentist approach are simply $1 - FDR$, whereas for the Bayesian approach, we compute the expected PPVs as $E_D(t^l)/N(D, a)$—that is, the estimated number of non-spurious arcs divided by the total number of arcs in the learned model. The curves do not extend to small values for expected PPV, because this region corresponds to graphs with many arcs which are computationally expensive to learn.

The Bayesian method produced curves that appear to vary slightly less across data sets than those of the frequentist method. We note that the variation in the frequentist results was not due to the use of too few permutations ($Q$) when estimating FDR. In particular, doubling the number of permutations did not substantially change the result.

Another observation is that the Bayesian curves tend to lie parallel to, and just below the idealized line. Thus the Bayesian approach tends to be slightly over-confident, attributing more belief in the given arcs than is warranted. In contrast, the frequentist curves tend to stay reasonably accurate for high PPVs, and then gradually peel away from the idealized curve, in a conservative manner, attributing increasingly less belief in the given arcs than is warranted. In the discussion section, we provide one possible explanation for this observation. We note that in real applications, the PPV range of interest is typically in the high end (where the frequentist approach is still accurate), because one does not want an abundance of false hypotheses to pursue.

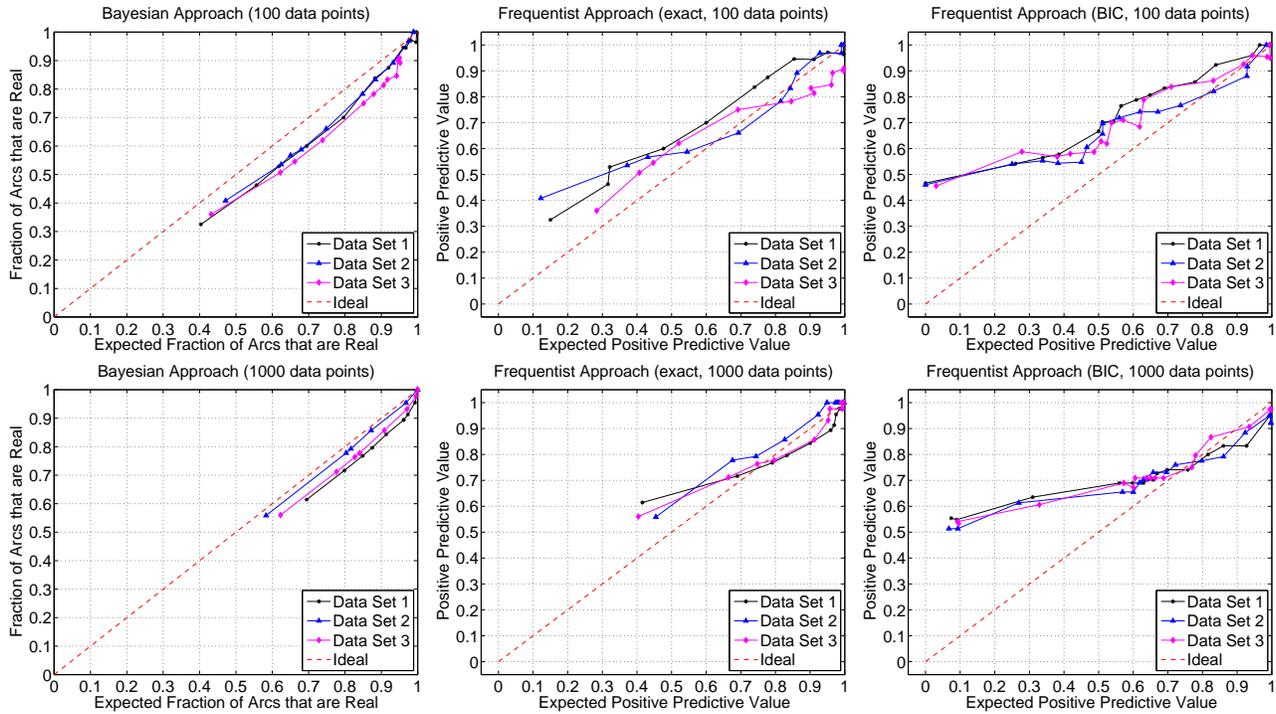

Figure 2: Comparison of the Bayesian and frequentist approaches. Actual versus estimated expected PPV are shown. The dashed line denotes the idealized curve, where actual and estimated expected PPV are equal.

**Sensitivity of Bayesian Method to the Prior**

Figure 3 shows the sensitivity of the Bayesian method to the prior hyperparameters $\kappa$ and $\alpha$. To generate the curves in the figure, we used the synthetic data sets from the previous section. To trace out curves across the full range of expected PPV, however, we used a structure-search method that was different from the one used in the previous section. In particular, we first computed marginal posterior probabilities of each arc hypothesis, again limiting our sum over all structures to those with parents sets of size five or less. We then used these marginal posterior probabilities to rank arc hypotheses and constructed a series of nested models, including in each model all arcs with marginal posteriors above some threshold.

We see that the curves are sensitive to the choice of hyperparameters. Furthermore, for the hyperparameters used in the previous section ('pred. optimal'), $E_D(t^l)$ remains relatively accurate across the entire range of expected PPV. Perhaps most interesting, however, is that the hyperparameters corresponding to the curve closest to ideal are not those that produce optimal out-of-sample predictions.

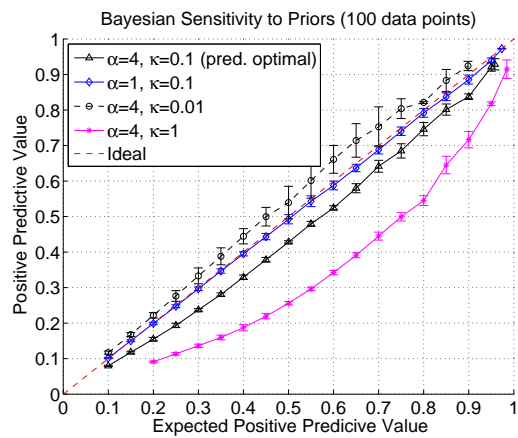

Figure 3: Sensitivity of the Bayesian approach to hyperparameters $\kappa$ and $\alpha$. Error bars denote one standard deviation above, and one below the mean from three data sets of size 100 (same data sets across all curves). The dashed line denotes the idealized curve.

**Synthetic HIV Data**

Next, we assessed the accuracy of our frequentist approach when applied to synthetic data resembling our real HIV data. Here, we could not evaluate the Bayesian approach

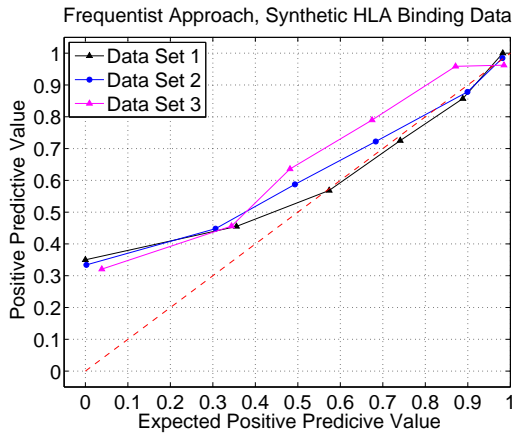

Figure 4: Estimated FDR for three sampled data sets from a single, synthetic, noisy-OR based HIV model, using the frequentist approach. The dashed line denotes the idealized curve.

as it was computationally infeasible.

When computing the FDR, we used a BIC-based greedy search (with no smoothing on the parameters). To compute the BIC score, we needed to find the maximum likelihood solution for noisy-OR nodes. Fortunately, this is a convex optimization problem [15].

Our real data consisted of observed interactions between 140 peptides and 102 patients, each with three to six HLA molecules spanning a set of 70 distinct HLA molecules. The model used to generate synthetic data was created with our greedy search algorithm on the real data with an estimated $FDR \approx 0.3$. Plots of actual versus expected PPV for three data sets generated from this model are shown in Figure 4. Estimates of PPV are quite accurate at the upper end (i.e., lower FDR), which is the region of interest for our problem and, as mentioned, the region of interest for many real problems.

**Results on Real Data**

Using the real peptide–HLA data, we found 168 peptide–HLA relationships at $FDR \approx 0.3$ among the possible $140 \times 70$ pairs. To validate our predictions on the real peptide–HLA data set, Frahm *et al.* (in submission) performed in vitro assays that specifically measured related HLA-peptide pairs. Ideally, all 168 pairs should have been evaluated, but this was too expensive. Consequently, they evaluated eight pairs for which the HLA-peptide association is biologically interesting (i.e., unlikely based on current understanding of peptide–HLA binding). All eight relationships were confirmed.

## 6 Discussion

We have investigated the use of a Bayesian and a frequentist approach to determining the number of non-spurious arcs in a learned DAG model with discrete variables and known variable ordering. Both methods take as input a model learned by an arbitrary method. In addition, the Bayesian method requires a prior distribution over all possible model structures and their parameters, whereas the frequentist method requires a specification of the method that produced the input model. Empirically, we have found that both methods produce fairly accurate estimates for lower values of FDR, a region that is typically of interest. We speculate, however, that the FDR approach, which is non-parametric, may be more robust to data which has not been generated from models in the same class as the learned models (*e.g.* CPT or noisy-OR distributions with parameter independence). Further investigation is needed to confirm this possibility.

In our empirical studies, we found that the frequentist estimation of FDR was overly conservative for larger values of FDR (smaller values of expected PPV). One explanation for this observation comes from the fact that we used null distributions derived from permutations of the data—that, is null distributions in which no arc hypotheses were true. To understand how this use of the null distribution may lead to conservative values for FDR, recall that arc hypotheses are not independent from one another so that any structure learning algorithm must take into account the presence of other arcs when deciding whether to include any particular arc. Consequently, when we apply our structure search algorithm to permutations of the data, arcs that are added later in the search have been built upon already spurious arcs (by construction) and therefore are not truly representative of spuriously generated arcs found with the real data, where spurious arcs are being added to mostly non-spurious arcs. Because a spurious arc does not tend to explain much of the data, it is more likely that another spurious arc could help explain the data than had the already present arcs been non-spurious. That is, when a true arc is already present, it is less likely that an additional spurious arc will help explain the data, because there is less explaining left to do. Thus, it seems reasonable that too many spurious arcs will be generated when estimating $E_0[F(\kappa)]$ using null distributions generated by permutation.

One possible approach to overcome this shortcoming would be to use a sequential approximation to $E_0[F]$, where instead of simply permuting the data and applying our search algorithm as we have done in our experiments, we instead recursively estimate $E_0[F(\kappa)]$. Specifically, we could start with a relatively low value of $\kappa = \kappa_0$ so that only a few arcs are generated using the real data, and estimate $E_0[F(\kappa_0)]$ as we have done. Then, to estimate $E_0[F(\kappa)]$ for a larger value of $\kappa$, $\kappa = \kappa_1$, we could (1) initialize structure search using the $\kappa_0$ structure and (2) continue structure search using null data generated from this same structure (using a parametric bootstrap when necessary). We could then use the estimate $E_0[F(\kappa_1)] \simeq E_0[F(\kappa_0)] + n_1$, where $n_1$ is the number of arcs added to the $\kappa_0$ structure. This recursive procedure could then be continued, increasing $\kappa$ on each iteration.

This extension deserves further investigation, although we note that the simple permutation approach is fairly accurate in regions of small FDR—the regions that are typically of interest to researchers.

Other work may include a more thorough study of how the FDR-based approach works for node parameterizations such as logistic regression and decision trees. Additionally, it would be interesting to evaluate the Bayesian and frequentist methods when used with search methods that do not assume a fixed variable ordering.

**Acknowledgments**

We thank the reviewers for their insightful comments. We also thank Nicole Frahm and Christian Brander for the HIV data which formed the basis of our synthetic peptide–HLA model, Chris Meek for help in finding the convexity results in [15], Carl Kadie for implementing software that was used for preliminary work with the peptide–HLA model and Moises Goldszmidt for useful discussions.